\pdfoutput=1
\documentclass[12pt]{elsarticle}
\usepackage[dvipsnames,svgnames,table]{xcolor}
\usepackage{graphicx} 
\usepackage{tikz}
\usepackage{pgfplots}
\pgfplotsset{compat=1.18}
\usepackage{standalone}
\usepackage{enumitem}
\usepackage{rotating}
\usepackage{booktabs}
\usepackage{tabularx}
\newcolumntype{R}{>{\raggedleft\arraybackslash}X}
\newcolumntype{L}{>{\raggedright\arraybackslash}X}
\newcolumntype{C}{>{\centering\arraybackslash}X}
\usepackage{threeparttable}

\usepackage[justification=centerfirst]{caption}
\usepackage[labelformat=parens]{subcaption}

\usepackage[hyphens]{url}
\usepackage{hyperref} 
\hypersetup{
  colorlinks   = true,    		
  urlcolor     = blue,    		
  linkcolor    = blue,    	    
  citecolor    = ForestGreen	    
}

\begin{document}

\begin{frontmatter}
\title{How to measure consumer's  inconsistency in sensory testing?}

\author[mate,krtk]{L\'aszló Sipos}\corref{cor1}
\ead{sipos.laszlo@uni-mate.hu, sipos.laszlo@krtk.hun-ren.hu}
\author[cor]{Kolos Csaba \'Agoston}
\ead{kolos.agoston@uni-corvinus.hu}
\author[krtk,cor]{P\'eter Bir\'o}
\ead{peter.biro@krtk.hun-ren.hu}
\author[sztaki,cor]{S\'andor Boz\'oki}
\ead{bozoki.sandor@sztaki.hun-ren.hu}
\author[sztaki,cor]{L\'aszl\'o Csat\'o}
\ead{laszlo.csato@sztaki.hun-ren.hu}

\cortext[cor1]{~Corresponding author}
\address[mate]{~Department of Postharvest, Commercial and Sensory Science, Institute of Food Science and Technology, Hungarian University of Agriculture and Life Sciences, Budapest, Hungary}
\address[krtk]{~HUN-REN Institute of Economics, Centre for Economic and Regional Studies (HUN-REN KRTK), Budapest, Hungary}
\address[cor]{~Department of Operations Research and Actuarial Sciences, Institute of Operations and Decision Sciences, Corvinus University of Budapest, Hungary}
\address[sztaki]{~HUN-REN Institute for Computer Science and Control (HUN-REN SZTAKI), Laboratory on Engineering and Management Intelligence, Research Group of Operations Research and Decision Systems, Budapest, Hungary}

\date{December 2023}

\begin{abstract}
\noindent
Standard methods, standard test conditions and the use of good sensory practices are key elements of sensory testing. However, while compliance assessment by trained and expert assessors is well developed, few information is available on testing consumer consistency. Therefore, we aim to answer the following research questions:
What type of metrics can be used to characterise (in)consistent evaluation?
How to rank assessors and attributes' evaluations based on (in)consistency?
Who can be considered an (in)consistent assessor?
What is the difference between consistent and inconsistent assessors' evaluations and use of scale?
What is the impact of consistent and inconsistent assessors' evaluations on the overall liking estimate?
The proposed detection of (in)consistency requires evaluations on two connected scales.
We reveal how assessors and attributes can be ranked according to the level of inconsistency, as well as how inconsistent assessors can be identified.
The suggested approach is illustrated by data from sensory tests of biscuits enriched with three pollens at different levels.
Future consumer tests are recommended to account for possible inconsistent evaluations.
\end{abstract}

\begin{keyword}
consistency \sep consumer sensory testing \sep JAR scale \sep liking score \sep product development
\end{keyword}
\end{frontmatter}

\section{Introduction}
Consumer testing is used in many areas of sensory research and food practice. The use of liking/acceptability feedback from target consumers is wide-ranging and includes comparing competing products, optimising product development, determining shelf-life, obtaining the liking of sensory attributes, assessing the impact of a product composition/packaging change on liking, etc. The main focus is to test whether or not there is a perceived difference in liking as a result of a treatment/change/modification. Consumer tests should be conducted with circumspection, taking into account many factors such as: purpose of the research, type of product, number of products, definition of target population (market product/new product, current users/potential users, preliminary results, specific segments), test location, expected accuracy of measurement, test procedure, product presentation plan for the chosen test procedure, specifications of the hypothesis to be tested (composition and size of the consumer sample) \citep{ISO_11136}.

In practice, there are a number of available consumer testing methods, from which the most appropriate method for the research objective should be chosen. There are two main groups of hedonic tests:
1) acceptability tests that measure the intensity of pleasure of consumption; and
2) preference tests that measure the order of preference of different products (pairwise comparison test, ranking).
Preference tests are relative and do not provide information on acceptability. In addition to the test method, it is also necessary to decide on the
location of the test: central location test (CLT), point of sale test (PST), home use test (HUT), home panel test (HPT), in-house test (IHT), which can be characterised by different advantages and disadvantages. The main advantage of tests carried out in a sensory laboratory is that they are based on good sensory practice and are carried out under controlled conditions. The advantages and disadvantages of different testing sites have already been discussed in the literature \citep{ISO_8589, ISO_11136, Boutrolle_etal2007, HastedMcEwanJenner2023, LawlessHeymann2010, Munoz2013, Park_etal2023, StoneBleibaumThomas2020}.

International standard methods, standard test conditions and the use of good sensory practices are key elements of sensory testing, too. The characteristics of a sensory expert include demonstrated sensory sensitivity, extensive training and experience, and the ability to provide consistent and repeatable evaluations \citep{ISO_5492}. Compliance assessment by trained and expert assessors is well developed: measurement of discrimination, agreement,  repeatability \citep{ISO_11132}, consonance \citep{AguirreHuerta-PachecoLopez2017, Dijksterhuis1995, KermitLengard2005}, scaling usage and scaling effect \citep{BrockhoffSchlichSkovgaard2015, PeltierVisalliSchlich2016, Romano_etal2008}, statistical methods and software for performance measurement of sensory panelists and panels \citep{Sipos_etal2021}, workflow of panel performance \citep{NaesBrockhoffTomic2011}, characteristics of trained panels \citep{Djekic_etal2021}, panel management \citep{Rogers2017}, panel and panelist performance of Temporal Dominance of Sensations \citep{Frascolla_etal2023, Lepage_etal2014, Meyners2011, Schlich2017}, sensory analysis programme in quality control \citep{ISO_20613, Munoz2013, PineauChabanetSchlich2007}.

In sensory testing, consistency refers on the one hand to a mechanical attribute perceived by mechanical tactile or visual receptors \citep{ISO_5492}. On the other hand, the consistency is one of the performance characteristics of a trained and expert assessor or panel, along with discrimination ability, repeatability ability, agreement and scale usage, which can be monitored and evaluated in one session or over time. According to the international standard for guidelines for the measurement of quantitative descriptive sensory panel performance the consistency of an assessor is inversely related to the standard deviation of the bias terms (difference between the assessor’s mean for the sample and the panel mean) calculated from each sample. Using ANOVA model, both panel consistency (session to session standard deviation with control sample) and internal consistency (interaction of assessor and session) can be determined \citep{ISO_11132}. Internal consistency test based on Cronbach's alpha is an accepted method, since it allows identifying attributes better understood by panelists, and it gives a ranking of panelists according to their consensus with the rest of the panel \citep{PintoFogliattoQannari2014}. Other multivariate methods are also used to measure the (in)consistency of panels, such as Wilk's lambda in MANOVA models \citep{FernandoSamitaPeiris2022}, or gravity centre and area/perimeter (GCAP) method \citep{Kollar-Hunek_etal2008}, and parallel factor analysis (PARAFAC) \citep{RomanoVestergaardKompany-ZarehBredie-2011}.

The problem of inconsistent consumer response is typically associated with questionnaire surveys in consumer research. It is a common goal when designing a questionnaire to filter out inconsistent consumer responses after the test, so the same questions are often asked in different ways or on different types of response scales. Obtaining unbiased and consistent study results depends heavily on the way the survey is administered and distributed to respondents and how respondents are sampled from the target population \citep{Brace_2018}. While there are many developed measures, indices and statistical methods for trained and expert sensory assessors, there are no developed statistical indicators for measuring consumer (in)consistency. The following research questions have been identified:
\begin{enumerate}
\item 
What type of metrics can be used to characterise (in)consistent evaluation?
\item 
How to rank assessors and the evaluations of attributes based on \linebreak (in)consistency?
\item 
Who can be considered an (in)consistent assessor?
\item 
What is the difference between consistent and inconsistent assessors' evaluation and use of scale?
\item
What is the impact of consistent and inconsistent assessors' evaluations on the overall liking estimate?
\end{enumerate}

\section{Material and Methods}

\subsection{Biscuit Preparation}

Biscuits were prepared according to the AACC-approved method 10–50D \citep{AACC_1980}. The control sample contained ingredients included in the standard. Ground bee pollens were used to substitute 2\%, 5\%, or 10\% of wheat flour (C control; RS2, RS5, RS10 rapeseed; SF2, SF5, SF10 sunflower; PH2, PH5, PH10 phacelia).

All ingredients were weighed in a plastic bowl with a precision of two decimal places, then mixed into a homogenous mass. The dough was sheeted to a thickness of 7mm. Biscuits with a diameter of 50mm were formed and baked in an electrically heated rotary oven (Gierre, Milano, Italy) for 10 minutes at 205$^{\circ}$C. The samples were cooled to room temperature and placed in sealable plastic bags for testing within one hour.

\subsection{Consumers’ preference tests}

Sensory tests were carried out in a sensory laboratory (Hungarian University of Agriculture and Life Sciences, Institute of Food Science and Technology, Department of Postharvest, Commercial and Sensory Science) that meet the standard requirements \citep{ISO_8589}. The work was conducted in accordance with The Code of Ethics of the World Medical Association (Declaration of Helsinki). The tests were carried out anonymously and on a voluntary basis. Participants gave informed consent via the statement ``I am aware that my responses are confidential, and I agree to participate in this experiment''; an affirmative reply was required to enter the test. They were able to withdraw from the experiment at any time without giving a reason. The products tested were safe for consumption. Before the test, participants were informed that the biscuits contained pollen and gluten, which may cause hypersensitivity reactions in sensitive individuals.
The tests were designed and implemented according to the international standard for consumer preference tests \citep{ISO_11136}. A total of 100 consumers participated. Inclusion criteria were as follows: willingness to participate in the experiment, does not have a history of food allergies or intolerances including sensitivity to beekeeping products. Each participant evaluated biscuits on two sessions to minimize fatigue. Products were coded with random, three-digit numbers starting with non-zero. 
Orders of biscuit samples were balanced.
As a taste neutralizer, mineral water with a neutral taste was provided for each participant.

Consumers evaluated the different samples on two different scales. Firstly, the biscuits were evaluated according to their liking, based on the following 10 attributes: darkness, overall odour, sweet odour, margarine odour, overall taste, sweet taste, margarine taste, hardness, crumbliness, and overall acceptance. A nine-category monotonic ascending hedonic response scale was used, where each category was associated with emoticons and descriptive terms (1 = extremely dislike, 2 = strongly dislike, 3 = moderately dislike, 4 = slightly dislike, 5 = neither like nor dislike, 6 = slightly like, 7 = moderately like, 8 = strongly like, 9 = extremely like). Secondly, the biscuits were evaluated according to their intensity values, including the following 9 attributes: darkness, overall odour, sweet odour, margarine odour, overall taste, sweet taste, margarine taste, hardness, crumbliness, on a 5-category just about right (JAR) scale ($-$2 = not enough at all, $-$1 = not enough, 0 = just about right, 1 = too much, 2 = far too much) \citep{ISO_4121}.

\section{Results} \label{Sec3}

This section presents our approach for measuring inconsistency and applies it to the biscuit dataset.
Tables~\ref{Table1} and \ref{Table2} show the means and standard deviations of the liking and JAR intensity scores for the 10 different types of products.


\begin{sidewaystable}
  \centering
  \caption{Descriptive statistics of liking scores}
\label{Table1}
\begin{threeparttable}
\rowcolors{1}{}{gray!20}
\resizebox{\textwidth}{!}{
    \begin{tabular}{lc ccc ccc ccc c} \toprule
        Attribute & C & RS2 & RS5 & RS10 & SF2 & SF5 & SF10 & PH2 & PH5 & PH10 & All  \\ \bottomrule
        Colour &\begin{tabular}{c}6.18\\(1.93) \end{tabular}&\begin{tabular}{c}6.36 \\ (1.43)  \end{tabular}&\begin{tabular}{c}6.92 \\ (1.69) \end{tabular}&\begin{tabular}{c}6.73 \\ (1.77) \end{tabular}&\begin{tabular}{c}6.81 \\ (1.66) \end{tabular}&\begin{tabular}{c}6.91 \\ (1.56) \end{tabular}&\begin{tabular}{c}6.65 \\ (1.86) \end{tabular}&\begin{tabular}{c}5.81 \\ (1.94) \end{tabular}&\begin{tabular}{c}4.73 \\(1.95)\end{tabular}&\begin{tabular}{c}4.66 \\ (2.28) \end{tabular}&\begin{tabular}{c}6.18 \\ (1.99) \end{tabular}\\ 
        Global odour &\begin{tabular}{c}5.87 \\ (1.60) \end{tabular} &\begin{tabular}{c}5.68 \\ (1.85) \end{tabular}&\begin{tabular}{c}5.56 \\ (2.00) \end{tabular}&\begin{tabular}{c}4.65 \\ (2.43) \end{tabular}&\begin{tabular}{c}6.18 \\ (1.51) \end{tabular}&\begin{tabular}{c}6.12 \\ (1.57) \end{tabular} &\begin{tabular}{c}5.81 \\ (2.18) \end{tabular} &\begin{tabular}{c}5.84 \\ (1.65) \end{tabular} &\begin{tabular}{c}5.61 \\ (1.52) \end{tabular} &\begin{tabular}{c}4.38 \\ (1.98) \end{tabular} &\begin{tabular}{c}5.57 \\ (1.93) \end{tabular}\\ 
        Sweet odour &\begin{tabular}{c}5.58 \\ (1.94) \end{tabular}&\begin{tabular}{c}5.51 \\ (1.90) \end{tabular}&\begin{tabular}{c}5.78 \\ (1.90) \end{tabular}&\begin{tabular}{c}4.59 \\ (2.23) \end{tabular}&\begin{tabular}{c}5.67 \\ (1.65) \end{tabular} &\begin{tabular}{c}5.79 \\ (1.71) \end{tabular} &\begin{tabular}{c}5.46 \\ (1.97) \end{tabular} &\begin{tabular}{c}5.46 \\ (1.78) \end{tabular} &\begin{tabular}{c}5.35 \\ (1.64) \end{tabular} &\begin{tabular}{c}4.20 \\ (1.86) \end{tabular} &\begin{tabular}{c}5.34 \\ (1.92) \end{tabular} \\ 
        Margarine odour &\begin{tabular}{c}5.38 \\ (1.25) \end{tabular}&\begin{tabular}{c}5.37 \\ (1.61) \end{tabular}&\begin{tabular}{c}5.43 \\ (1.53) \end{tabular}&\begin{tabular}{c}4.52 \\ (1.94) \end{tabular}&\begin{tabular}{c}5.85 \\ (1.42) \end{tabular}&\begin{tabular}{c}5.42 \\ (1.49) \end{tabular} &\begin{tabular}{c}5.49 \\ (1.55) \end{tabular} &\begin{tabular}{c}5.44 \\ (1.54) \end{tabular} &\begin{tabular}{c}5.40 \\ (1.63) \end{tabular} &\begin{tabular}{c}4.46 \\ (1.68) \end{tabular} &\begin{tabular}{c}5.28 \\ (1.63) \end{tabular}\\ 
        Global taste &\begin{tabular}{c}6.50 \\ (1.76) \end{tabular}&\begin{tabular}{c}5.76 \\ (2.16) \end{tabular}&\begin{tabular}{c}5.82 \\ (1.85) \end{tabular}&\begin{tabular}{c}4.63 \\ (2.58) \end{tabular} &\begin{tabular}{c}6.96 \\ (1.36) \end{tabular}&\begin{tabular}{c}6.89 \\ (1.59) \end{tabular} &\begin{tabular}{c}6.10 \\ (2.16) \end{tabular} &\begin{tabular}{c}6.71 \\ (1.82) \end{tabular} &\begin{tabular}{c}6.63 \\ (1.59) \end{tabular} &\begin{tabular}{c}4.11 \\ (2.42) \end{tabular} &\begin{tabular}{c}6.01 \\ (2.14) \end{tabular}\\ 
        Sweet taste &\begin{tabular}{c}6.57 \\ (1.58) \end{tabular}&\begin{tabular}{c}6.25 \\ (1.88) \end{tabular}&\begin{tabular}{c}6.38 \\ (1.80) \end{tabular}&\begin{tabular}{c}5.23 \\ (2.40) \end{tabular}&\begin{tabular}{c}6.94 \\ (1.39) \end{tabular} &\begin{tabular}{c}6.78 \\ (1.62) \end{tabular} &\begin{tabular}{c}6.03 \\ (1.91) \end{tabular} &\begin{tabular}{c}6.50 \\ (1.70) \end{tabular} &\begin{tabular}{c}6.60 \\ (1.54) \end{tabular} &\begin{tabular}{c}4.59 \\ (2.35) \end{tabular} &\begin{tabular}{c}6.19 \\ (1.97) \end{tabular}\\ 
        Margarine taste &\begin{tabular}{c}5.83 \\ (1.83) \end{tabular}&\begin{tabular}{c}5.36 \\ (1.76) \end{tabular}&\begin{tabular}{c}5.35 \\ (1.65) \end{tabular}&\begin{tabular}{c}4.9 \\ (2.04) \end{tabular} &\begin{tabular}{c}6.36 \\ (1.51) \end{tabular} &\begin{tabular}{c}6.16 \\ (1.45) \end{tabular} &\begin{tabular}{c}5.55 \\ (1.87) \end{tabular} &\begin{tabular}{c}6.15 \\ (1.60) \end{tabular} &\begin{tabular}{c}5.72 \\ (1.90) \end{tabular} &\begin{tabular}{c}4.30 \\ (1.88) \end{tabular} &\begin{tabular}{c}5.57 \\ (1.85) \end{tabular}\\ 
        Hardness &\begin{tabular}{c}4.92 \\ (2.27) \end{tabular} &\begin{tabular}{c}6.02 \\ (2.00) \end{tabular}&\begin{tabular}{c}4.98 \\ (2.25) \end{tabular}&\begin{tabular}{c}4.56 \\ (2.57) \end{tabular}&\begin{tabular}{c}6.62 \\ (2.00) \end{tabular}&\begin{tabular}{c}5.62 \\ (2.21) \end{tabular}&\begin{tabular}{c}4.92 \\ (2.48) \end{tabular}&\begin{tabular}{c}5.58 \\ (2.38) \end{tabular}&\begin{tabular}{c}4.39 \\ (2.38)\end{tabular}&\begin{tabular}{c}3.58 \\ (2.54) \end{tabular}&\begin{tabular}{c}5.12 \\ (2.45) \end{tabular}\\ 
        Crumbleness &\begin{tabular}{c}5.61 \\ (2.05) \end{tabular}&\begin{tabular}{c}6.26 \\ (1.91) \end{tabular}&\begin{tabular}{c}5.70 \\ (2.02) \end{tabular}&\begin{tabular}{c}5.36 \\ (2.32) \end{tabular}&\begin{tabular}{c}6.39 \\ (1.88) \end{tabular} &\begin{tabular}{c}6.18 \\ (1.97) \end{tabular} &\begin{tabular}{c}5.57 \\ (2.02) \end{tabular} &\begin{tabular}{c}5.60 \\ (2.16) \end{tabular} &\begin{tabular}{c}5.39 \\ (2.07) \end{tabular} &\begin{tabular}{c}4.87 \\ (2.34) \end{tabular} &\begin{tabular}{c}5.69 \\ (2.12) \end{tabular}\\ \toprule
        Global liking &\begin{tabular}{c}6.29 \\ (1.79) \end{tabular}&\begin{tabular}{c}5.94 \\ (2.21) \end{tabular}&\begin{tabular}{c}5.99 \\ (1.93) \end{tabular}&\begin{tabular}{c}4.69 \\ (2.59) \end{tabular}&\begin{tabular}{c}6.93 \\ (1.63) \end{tabular} &\begin{tabular}{c}6.58 \\ (1.82) \end{tabular} &\begin{tabular}{c}5.80 \\ (2.09) \end{tabular} &\begin{tabular}{c}6.53 \\ (1.71) \end{tabular}&\begin{tabular}{c}5.81 \\ (1.96) \end{tabular} &\begin{tabular}{c}3.83 \\ (2.34) \end{tabular} &\begin{tabular}{c}5.84 \\ (2.20) \end{tabular}\\ \bottomrule
    \end{tabular}
}
\begin{tablenotes} \scriptsize
    \item \emph{Notes:} Means are at the top, standard deviations are at the bottom in parenthesis. 
    \item C: control.
    \item RS2, RS5, RS10: rapeseed (2\%, 5\%, 10\%).
    \item SF2, SF5, SF10: sunflower (2\%, 5\%, 10\%).
    \item PH2, PH5, PH10: phacelia (2\%, 5\%, 10\%).
\end{tablenotes}
\end{threeparttable}
\end{sidewaystable}

\begin{sidewaystable}
  \caption{Descriptive statistics of JAR scores}
\label{Table2}
\begin{threeparttable}
\rowcolors{1}{}{gray!20}
\resizebox{\textwidth}{!}{
    \begin{tabular}{lc ccc ccc ccc c} \toprule
        Attribute & C & RS2 & RS5 & RS10 & SF2 & SF5 & SF10 & PH2 & PH5 & PH10 & All  \\ \bottomrule
        Colour          &\begin{tabular}{c}{\it $-$0.83} \\ (0.84) \end{tabular} &\begin{tabular}{c}{\it $-$0.69} \\ (0.75) \end{tabular} &\begin{tabular}{c}{\it $-$0.29} \\ (0.48) \end{tabular} &\begin{tabular}{c}{\it  0.54} \\ (0.74) \end{tabular} &\begin{tabular}{c}{\it $-$0.21} \\ (0.62) \end{tabular} &\begin{tabular}{c}{\it  0.40} \\ (0.55) \end{tabular} &\begin{tabular}{c}{\it  0.51} \\ (0.69) \end{tabular} &\begin{tabular}{c}{    $-$0.13} \\ (0.92) \end{tabular} &\begin{tabular}{c}{\it  0.90} \\ (1.05) \end{tabular} &\begin{tabular}{c}{\it  1.56} \\ (0.72) \end{tabular} &\begin{tabular}{c}{\it  0.18} \\ (1.03) \end{tabular}\\ 
        Global odour    &\begin{tabular}{c}{\it $-$0.74} \\ (0.73) \end{tabular} &\begin{tabular}{c}{    $-$0.16} \\ (0.94) \end{tabular} &\begin{tabular}{c}{    $-$0.08} \\ (0.75) \end{tabular} &\begin{tabular}{c}{\it  0.80} \\ (1.02) \end{tabular} &\begin{tabular}{c}{\it $-$0.49} \\ (0.81) \end{tabular} &\begin{tabular}{c}{    $-$0.04} \\ (0.85) \end{tabular} &\begin{tabular}{c}{     0.18} \\ (1.00) \end{tabular} &\begin{tabular}{c}{\it $-$0.23} \\ (0.81) \end{tabular} &\begin{tabular}{c}{    $-$0.10} \\ (0.88) \end{tabular} &\begin{tabular}{c}{\it  0.80} \\ (1.24) \end{tabular} &\begin{tabular}{c}{    $-$0.01} \\ (1.03) \end{tabular}\\ 
        Sweet odour     &\begin{tabular}{c}{\it $-$0.86} \\ (0.80) \end{tabular} &\begin{tabular}{c}{\it $-$0.43} \\ (0.89) \end{tabular}  &\begin{tabular}{c}{\it $-$0.39} \\ (0.82) \end{tabular}  &\begin{tabular}{c}{    $-$0.11} \\ (1.19) \end{tabular}  &\begin{tabular}{c}{\it $-$0.54} \\ (0.74) \end{tabular}  &\begin{tabular}{c}{\it $-$0.36} \\ (0.84) \end{tabular}  &\begin{tabular}{c}{\it $-$0.60} \\ (0.93) \end{tabular}  &\begin{tabular}{c}{\it $-$0.51} \\ (0.90) \end{tabular}  &\begin{tabular}{c}{\it $-$0.61} \\ (0.79) \end{tabular}  &\begin{tabular}{c}{\it $-$0.86} \\ (1.18) \end{tabular}  &\begin{tabular}{c}{\it $-$0.53} \\ (0.94) \end{tabular} \\ 
        Margarine odour &\begin{tabular}{c}{\it $-$0.59} \\ (0.75) \end{tabular} &\begin{tabular}{c}{    $-$0.16} \\ (0.90) \end{tabular} &\begin{tabular}{c}{\it $-$0.44} \\ (0.84) \end{tabular} &\begin{tabular}{c}{    $-$0.18} \\ (1.17) \end{tabular} &\begin{tabular}{c}{\it $-$0.29} \\ (0.69) \end{tabular} &\begin{tabular}{c}{\it $-$0.16} \\ (0.80) \end{tabular} &\begin{tabular}{c}{\it $-$0.37} \\ (0.99) \end{tabular} &\begin{tabular}{c}{\it $-$0.32} \\ (0.85) \end{tabular} &\begin{tabular}{c}{\it $-$0.42} \\ (0.94) \end{tabular} &\begin{tabular}{c}{\it $-$0.48} \\ (1.26) \end{tabular} &\begin{tabular}{c}{\it $-$0.34} \\ (0.94) \end{tabular}\\ 
        Global taste    &\begin{tabular}{c}{\it $-$0.16} \\ (0.80) \end{tabular}&\begin{tabular}{c}{\it  0.27} \\ (0.85) \end{tabular} &\begin{tabular}{c}{\it  0.18} \\ (0.70) \end{tabular} &\begin{tabular}{c}{\it  0.94} \\ (1.04) \end{tabular} &\begin{tabular}{c}{\it $-$0.23} \\ (0.62) \end{tabular} &\begin{tabular}{c}{\it  0.17} \\ (0.71) \end{tabular} &\begin{tabular}{c}{\it  0.35} \\ (0.95) \end{tabular} &\begin{tabular}{c}{   $-$0.11} \\ (0.60) \end{tabular} &\begin{tabular}{c}{     0.05} \\ (0.73) \end{tabular} &\begin{tabular}{c}{\it  0.85} \\ (1.27) \end{tabular} &\begin{tabular}{c}{\it  0.23} \\ (0.93) \end{tabular}\\ 
        Sweet taste     &\begin{tabular}{c}{    $-$0.09} \\ (0.81) \end{tabular} &\begin{tabular}{c}{\it  0.24} \\ (0.88) \end{tabular}  &\begin{tabular}{c}{\it  0.23} \\ (0.74) \end{tabular}  &\begin{tabular}{c}{     0.18}\\ (1.22) \end{tabular}  &\begin{tabular}{c}{    $-$0.10} \\ (0.61) \end{tabular}  &\begin{tabular}{c}{    $-$0.06} \\ (0.75) \end{tabular}  &\begin{tabular}{c}{     0.05} \\ (0.97) \end{tabular}  &\begin{tabular}{c}{    $-$0.01} \\ (0.73) \end{tabular}  &\begin{tabular}{c}{\it $-$0.25} \\ (0.74) \end{tabular}  &\begin{tabular}{c}{\it $-$0.72} \\ (1.21) \end{tabular}  &\begin{tabular}{c}{    $-$0.05} \\ (0.92) \end{tabular} \\ 
        Margarine taste &\begin{tabular}{c}{    $-$0.16} \\ (0.92) \end{tabular} &\begin{tabular}{c}{    $-$0.10} \\ (0.89) \end{tabular} &\begin{tabular}{c}{    $-$0.14} \\ (0.79) \end{tabular} &\begin{tabular}{c}{    $-$0.03} \\ (1.11) \end{tabular} &\begin{tabular}{c}{\it $-$0.15} \\ (0.69) \end{tabular} &\begin{tabular}{c}{    $-$0.08} \\ (0.77) \end{tabular} &\begin{tabular}{c}{    $-$0.10} \\ (0.93) \end{tabular} &\begin{tabular}{c}{    $-$0.10} \\ (0.70) \end{tabular} &\begin{tabular}{c}{\it $-$0.19} \\ (0.80) \end{tabular} &\begin{tabular}{c}{\it $-$0.51} \\ (1.13) \end{tabular} &\begin{tabular}{c}{\it $-$0.16} \\ (0.89) \end{tabular}\\ 
        Hardness        &\begin{tabular}{c}{\it  0.42} \\ (0.94) \end{tabular} &\begin{tabular}{c}{    $-$0.15} \\ (0.78) \end{tabular} &\begin{tabular}{c}{\it  0.53} \\ (0.76) \end{tabular} &\begin{tabular}{c}{\it  1.02} \\ (0.89) \end{tabular} &\begin{tabular}{c}{     0.01} \\ (0.67) \end{tabular} &\begin{tabular}{c}{\it  0.67} \\ (0.73) \end{tabular} &\begin{tabular}{c}{\it  0.88} \\ (0.86) \end{tabular} &\begin{tabular}{c}{\it 0.53} \\ (0.87) \end{tabular} &\begin{tabular}{c}{\it  0.88} \\ (0.87) \end{tabular} &\begin{tabular}{c}{\it  1.59} \\ (0.79) \end{tabular} &\begin{tabular}{c}{\it  0.64} \\ (0.94) \end{tabular}\\ 
        Crumbleness     &\begin{tabular}{c}{     0.03} \\ (0.80) \end{tabular}&\begin{tabular}{c}{    $-$0.06} \\ (0.80) \end{tabular} &\begin{tabular}{c}{\it  0.25} \\ (0.81) \end{tabular} &\begin{tabular}{c}{\it  0.40} \\ (0.92) \end{tabular} &\begin{tabular}{c}{\it  0.21} \\ (0.80) \end{tabular} &\begin{tabular}{c}{\it  0.33} \\ (0.80) \end{tabular} &\begin{tabular}{c}{\it  0.39} \\ (0.74) \end{tabular} &\begin{tabular}{c}{\it  0.28} \\ (0.74) \end{tabular} &\begin{tabular}{c}{\it  0.44} \\ (0.91) \end{tabular} &\begin{tabular}{c}{\it  0.27} \\ (1.14) \end{tabular} &\begin{tabular}{c}{\it  0.25} \\ (0.99) \end{tabular}\\ \toprule
    \end{tabular}
}
\begin{tablenotes} \scriptsize
    \item \emph{Notes:} Means are at the top, standarda deviations are at the bottom in parenthesis. 
    \item C: control.
    \item RS2, RS5, RS10: rapeseed (2\%, 5\%, 10\%).
    \item SF2, SF5, SF10: sunflower (2\%, 5\%, 10\%).
    \item PH2, PH5, PH10: phacelia (2\%, 5\%, 10\%).
    \item Numbers in \emph{italics} are significantly different from zero at 5\%.
\end{tablenotes}
\end{threeparttable}
\end{sidewaystable}

\subsection{The relationship between liking and JAR scores}

Based on the definitions of the liking and JAR intensity scales, the highest liking score (extremely liked, 9) is expected to be associated with the optimal level of intensity (just about right, 0).
Being less strict, a reasonable assumption for the consistency of evaluations is that a higher liking score implies a lower absolute JAR intensity score, and vice versa. 


\begin{table}[t!]
    \centering
    \caption{Contingency table of all liking and JAR scores}
    \label{Table3}
\begin{subtable}{\textwidth}
    \centering
    \caption{Numbers of evaluations}
    \label{Table3a}
    \rowcolors{1}{gray!20}{}
    \begin{tabularx}{0.6\textwidth}{lCCCCC} \toprule \hiderowcolors
 Liking  &  \multicolumn{5}{c}{JAR score} \\
 score   &  $-$2 &  $-$1  &  0  &  1  &  2 \\ \bottomrule \showrowcolors
1&	77&	39&	37&	43&	131\\
2&	92&	69&	69&	88&	145\\
3&	77&	142&	112&	163&	131\\
4&	83&	215&	240&	274&	114\\
5&	135&	477&	829&	270&	81\\
6&	57&	358&	625&	308&	54\\
7&	52&	304&	824&	304&	43\\
8&	22&	184&	943&	202&	45\\
9&	3&	36&	432&	46&	25\\ \toprule
    \end{tabularx}
\end{subtable}

\vspace{0.25cm}
\begin{subtable}{\textwidth}
    \centering
    \caption{Proportional distribution of JAR scores for each liking score}
    \label{Table3b}
    \rowcolors{1}{gray!20}{}
    \begin{tabularx}{0.8\textwidth}{lCCCCC} \toprule \hiderowcolors
 Liking  &  \multicolumn{5}{c}{JAR score} \\
 score   &  $-$2 &  $-$1  &  0  &  1  &  2 \\ \bottomrule \showrowcolors
1&  23.55\% &	11.93\% &	11.31\% &	13.15\% &	40.06\% \\
2&  19.87\% &	14.90\% &	14.90\% &	19.01\% &	31.32\% \\
3&  12.32\% &	22.72\% &	17.92\% &	26.08\% &	20.96\% \\
4&  \textcolor{white}{0}8.96\% &	23.22\% &	25.92\% &	29.59\% &	12.31\% \\
5&  \textcolor{gray!20}{0}7.53\% &	26.62\% &	46.26\% &	15.07\% &	\textcolor{gray!20}{0}4.52\% \\
6&  \textcolor{white}{0}4.07\% &	25.53\% &	44.58\% &	21.97\% &	\textcolor{white}{0}3.85\% \\
7&  \textcolor{gray!20}{0}3.41\% &	19.91\% &	53.96\% &	19.91\% &	\textcolor{gray!20}{0}2.82\% \\
8&  \textcolor{white}{0}1.58\% &	13.18\% &	67.55\% &	14.47\% &	\textcolor{white}{0}3.22\% \\
9&  \textcolor{gray!20}{0}0.55\% &	\textcolor{gray!20}{0}6.64\% &	79.70\% &	\textcolor{gray!20}{0}8.49\% &	\textcolor{gray!20}{0}4.61\% \\ \toprule
    \end{tabularx}
\end{subtable}
\end{table}

Table~\ref{Table3a} shows the contingency table of the liking scores and JAR intensity values when all evaluations are considered. The general picture is encouraging: for low liking scores, JAR scores of $-2$ or $+2$ are dominant and vice versa; for high liking scores, JAR score of $0$ is dominant. On the other hand, the lowest possible liking score (1) coincides with an optimal JAR intensity level (JAR score = 0) in 37 cases out of the total 9000. Similarly, in 28 ($25+3$) cases, the highest liking score (9) comes with a JAR score of $-2$ or $+2$, which indicates suspicious and possible unreliable opinions. These observations strengthen our idea that the issue of (in)consistency is worth analyzing deeper.

Table~\ref{Table3b} reports the normalized contingency table, where the relative frequencies sum to 1 in each row. This makes the distributions of JAR scores for different liking scores comparable. For example, for the JAR value of $-2$, the nominal frequencies corresponding to liking scores 1 and 2 is 77-92, which is explained by the fact that liking 2 was given more frequently. The normalized values 23.55\% and 19.87\% show the relation better: a JAR score of $-2$ has a higher probability to occur if the liking score is 1 rather than 2.
Unsurprisingly, the normalized frequencies are decreasing as a function of the liking score for the JAR value $-$2. However, this pattern does not hold for the JAR value $+$2 if the liking score is relatively high. Finally, the normalized frequencies are almost increasing for JAR$=0$, except for liking values 5 and 6. 

\subsection{Identifying inconsistent evaluations}

Given an attribute and a biscuit sample, we consider a pair of evaluations on liking and JAR scales \emph{consistent} if a higher liking score is associated with a smaller \emph{absolute} JAR score. The Kendall $\tau$ correlation coefficient, one of the widely used rank correlation coefficients, seems to be an appropriate  measure for this relationship. However, the original version of Kendall $\tau$ does not allow for ties, but ties are characteristic in these applications.

The modified versions of Kendall $\tau_A$, $\tau_B$ and $\tau_c$ allow ties but differ from each other in the way of handling them. Since the frequency of ties is quite high due to the narrow JAR scale, and the corresponding contingency table is non-squared (nine liking scores vs.\ three absolute JAR intensity scores), Kendall $\tau_C$---also known as Stuart $\tau_C$---seems to be the best option. Its range is the interval $[-1,+1]$, where $-1$ indicates perfect correlation between the two evaluations, that is, a higher liking score is always paired to a lower JAR intensity score, and vice versa. 
 
Assume that there are $n$ observations and a pair of values (liking$_i$, JAR$_i$) for each observation ($i=1,\ldots,n$). Furthermore:
\begin{itemize}[label=--]
    \item
Let $n_c$ denote the number of ``concordant'' pairs $(i,j)$, for which either liking$_i >$ liking$_j$ and $|$JAR$_i| < |$JAR$_j|$; or liking$_i <$ liking$_j$ and $|$JAR$_i| > |$JAR$_j|$ hold, where $|a|$ is the absolute value of $a$;

    \item
Let $n_d$ denote the number of ``discordant'' pairs $(i,j)$, for which either liking$_i >$ liking$_j$ and $|$JAR$_i| < |$JAR$_j|$; or liking$_i <$ liking$_j$ and $|$JAR$_i| > |$JAR$_j|$ hold;

    \item
Let $m$ denote the minimum of columns and rows in the crosstable.
\end{itemize}

The definition of Kendall $\tau_C$ is:
\[
\tau_C=\frac{n_c-n_d}{n^2(m-1)/m}.
\]

The Kendall $\tau_C$ values have been calculated using the {\texttt StuartTauC} function in package {\texttt DescTools} in {\texttt R}.

A perfectly consistent assessor, with Kendall $\tau_C = -1$, may give pairs of evaluations such as (9,0), (1,$-2$),  (5,$+1$), where the first coordinate is on the liking scale and the second one is on the JAR intensity scale. 
A totally inconsistent assessor, with Kendall $\tau_C = +1$, may give completely contradictory evaluation-pairs such as (9,$-2$), (7,$+1$), (1,0). 
Random evaluations would lead to Kendall $\tau_C = 0$ in expected value. Thus, if Kendall $\tau_C$ is closer to $-$1, a higher liking score is more likely to be associated with a lower absolute JAR value.

Consequently, the Kendall $\tau_C$ coefficient can be used to quantify the level of inconsistency. The coefficient should be negative for a consistent assessor---and it can be tested whether Kendall $\tau_C$ is significantly smaller than 0. This makes it possible to identify assessors whose evaluations seem to be random (it cannot be rejected that Kendall $\tau_C$ is zero), as well as assessors whose evaluations are worse than random (Kendall $\tau_C$ coefficient is significantly positive), which indicates a possible misunderstanding of the task.

\begin{table}[t!]
    \centering
    \caption{Contingency tables of the least and the most inconsistent assessors}
    \label{Table4}
\begin{subtable}{\textwidth}
    \centering
    \caption{The least inconsistent assessor (Kendall $\tau_C = -0.73$)}
    \label{Table4a}
    \rowcolors{1}{gray!20}{}
    \begin{tabularx}{0.6\textwidth}{lCCCCC} \toprule \hiderowcolors
 Liking  &  \multicolumn{5}{c}{JAR score} \\
 score   &  $-$2 &  $-$1  &  0  &  1  &  2 \\ \bottomrule \showrowcolors
1&  1&  1&  0&  0& 11\\
2&  3&  1&  0&  0&  3\\
3&  0&  0&  0&  0&  1\\
4&  0&  2&  0&  4&  3\\
5&  1&  4&  9&  0&  3\\
6&  0&  0&  8&  2&  0\\
7&  0&  1& 10&  4&  0\\
8&  0&  0&  6&  2&  0\\
9&  0&  0& 10&  0&  0\\ \toprule
    \end{tabularx}
\end{subtable}

\vspace{0.25cm}
\begin{subtable}{\textwidth}
    \centering
    \caption{The most inconsistent assessor (Kendall $\tau_C = 0.08$)}
    \label{Table4b}
    \rowcolors{1}{gray!20}{}
    \begin{tabularx}{0.6\textwidth}{lCCCCC} \toprule \hiderowcolors
 Liking  &  \multicolumn{5}{c}{JAR score} \\
 score   &  $-$2 &  $-$1  &  0  &  1  &  2 \\ \bottomrule \showrowcolors
1&  0&  0&  0&  0&  2\\
2&  0&  0&  8&  0&  0\\
3&  0&  1&  3&  0&  0\\
4&  0&  0&  1&  1&  2\\
5&  0&  0& 29&  3&  0\\
6&  0&  0& 11&  4&  0\\
7&  1&  1&  8&  4&  0\\
8&  1&  0&  7&  2&  0\\
9&  0&  1&  0&  0&  0\\ \toprule
    \end{tabularx}
\end{subtable}
\end{table}

Table~\ref{Table4} shows the least and the most inconsistent assessors in our sample. In particular, Table~\ref{Table4a} is the contingency table of the least inconsistent assessor (who has the smallest Kendall $\tau_C = -0.73$), while Table~\ref{Table4b} corresponds to the least consistent assessor (who has the smallest Kendall $\tau_C = 0.08$).

\begin{figure}[t!]
\center
\begin{tikzpicture}


\draw[fill=red!25] (0.5,    -2) circle  (0.04166667);
\draw[fill=red!25] (0.5,    -1) circle  (0.04166667);
\draw[fill=blue!25] (0.5,    2) circle  (0.1381927);


\draw[fill=red!25] (1,    -2) circle  (0.07216878);
\draw[fill=red!25] (1,    -1) circle  (0.04166667);
\draw[fill=blue!25] (1,    2) circle  (0.07216878);

\draw[fill=blue!25] (1.5,    2) circle  (0.04166667);


\draw[fill=red!25] (2,    -1) circle  (0.05892557);
\draw[fill=blue!25] (2,    1) circle  (0.08333333);
\draw[fill=blue!25] (2,    2) circle  (0.07216878);


\draw[fill=red!25] (2.5,    -2) circle  (0.04166667);
\draw[fill=red!25] (2.5,    -1) circle  (0.08333333);
\draw[fill=blue!25] (2.5,    0) circle  (0.125);
\draw[fill=blue!25] (2.5,    2) circle  (0.07216878);

\draw[fill=blue!25] (3,    0) circle  ( 0.1178511);
\draw[fill=blue!25] (3,    1) circle  (0.05892557);


\draw[fill=red!25] (3.5,    -1) circle  ( 0.04166667);
\draw[fill=blue!25] (3.5,    0) circle  (0.1317616);
\draw[fill=blue!25] (3.5,    1) circle  (0.08333333);

\draw[fill=blue!25] (4,    0) circle  (0.1020621);
\draw[fill=blue!25] (4,    1) circle  (0.05892557);

\draw[fill=blue!25] (4.5,    0) circle  (0.1317616);

\draw[->,very thick](-0.1,0) -- (5,0);

\node at (6,0) {\scriptsize{liking score}} ;

\draw[->, very thick](0,-2.5) -- (0,2.5);

\node[rotate=90] at (-1,0) {\scriptsize{JAR score}} ;

\draw[very thick](0.5,-0.1) -- (0.5,0.1);
\node at (0.5,-0.5){\scriptsize{1}};
\draw[very thick](1,-0.1) -- (1,0.1);
\node at (1,-0.5){\scriptsize{2}};
\draw[very thick](1.5,-0.1) -- (1.5,0.1);
\node at (1.5,-0.5){\scriptsize{3}};
\draw[very thick](2,-0.1) -- (2,0.1);
\node at (2,-0.5){\scriptsize{4}};
\draw[very thick](2.5,-0.1) -- (2.5,0.1);
\node at (2.5,-0.5){\scriptsize{5}};
\draw[very thick](3,-0.1) -- (3,0.1);
\node at (3,-0.5){\scriptsize{6}};
\draw[very thick](3.5,-0.1) -- (3.5,0.1);
\node at (3.5,-0.5){\scriptsize{7}};
\draw[very thick](4,-0.1) -- (4,0.1);
\node at (4,-0.5){\scriptsize{8}};
\draw[very thick](4.5,-0.1) -- (4.5,0.1);
\node at (4.5,-0.5){\scriptsize{9}};

\draw[very thick](-0.1,1) -- (0.1,1);
\node at (-0.5,1){\scriptsize{1}};
\draw[very thick](-0.1,2) -- (0.1,2);
\node at (-0.5,2){\scriptsize{2}};
\draw[very thick](-0.1,0) -- (0.1,0);
\node at (-0.5,0){\scriptsize{0}};
\draw[very thick](-0.1,-1) -- (0.1,-1);
\node at (-0.5,-1){\scriptsize{$-1$}};
\draw[very thick](-0.1,-2) -- (0.1,-2);
\node at (-0.5,-2){\scriptsize{$-2$}};

\end{tikzpicture}
\caption{JAR and liking scores of the least inconsistent assessor (see also Table~\ref{Table4a}).}
\label{Fig1}
\end{figure}
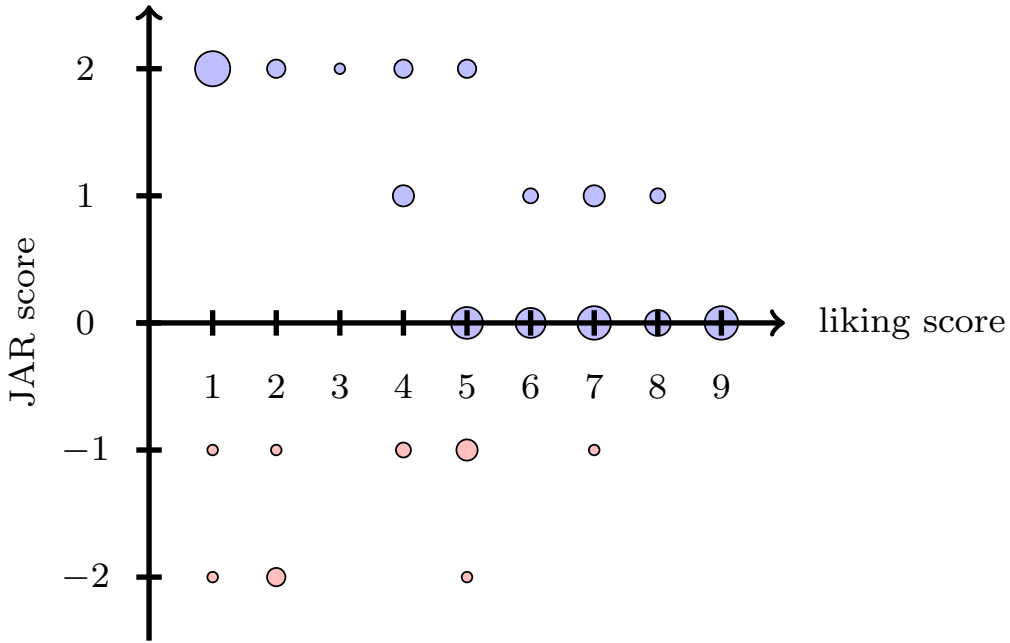

Table~\ref{Table4a} is visualized in Figure~\ref{Fig1}. Circle sizes are proportional to the frequencies of liking-JAR score pairs. 
The hypothetical perfectly consistent assessor, who gives (9,0), (1,$-2$), (1,$+2$),  (5,$-1$) and (5,$+1$), would result in a $>$ shape. The distribution of circles of the least inconsistent assessor shows a similar pattern.

Note that the intermediate liking score 5 frequently (29 times) occurs together with a JAR score of 0, representing an ideal level, in Table~\ref{Table4b}. 
This phenomenon can be observed  not only for the most inconsistent assessor, but also for many other ones. A possible explanation is a misunderstanding or misinterpretation of the two scales.

This phenomenon calls attention to a potential drawback of the Kendall $\tau_C$ rank correlation coefficient: neither concordant, nor disconcordant pairs can be identified if the liking or JAR values are equal. In an extreme case, if an assessor has only given JAR values of 0, then $\tau_C = 0$, independently of the liking values.

\subsection{The consistency of assessors' evaluations}

Consistency can be analysed with respect to both the assessors and the attributes, which leads to the rankings of assessors as well as attributes according to the proposed Kendall $tau_C$. 
Rank leaders' pairs of evaluations satisfy consistency the most frequently and to the highest degree. This way the definition of consistency can be applied to assessors or attributes.

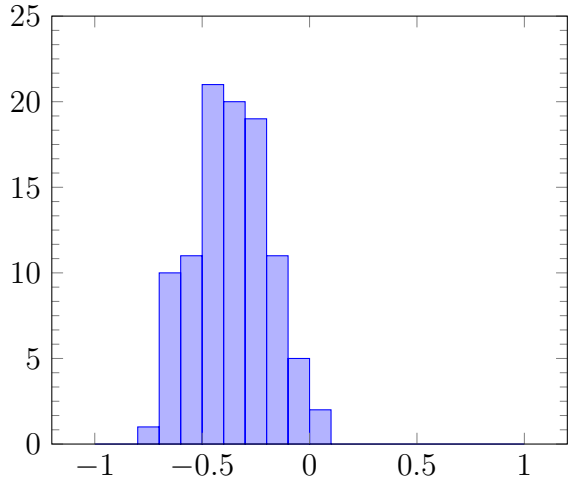
\begin{figure}[t!]
    \centering
\begin{tikzpicture}
\begin{axis}[
    ymin=0, ymax=25,
    minor y tick num = 5,
    area style,
    ]
\addplot+[ybar interval,mark=no] plot coordinates {
(-1, 0)
(-0.9, 0)
(-0.8, 1)
(-0.7, 10)
(-0.6, 11)
(-0.5, 21)
(-0.4, 20)
(-0.3, 19)
(-0.2, 11)
(-0.1, 5)
(0, 2)
(0.1, 0)
(0.2, 0)
(0.3, 0)
(0.4, 0)
(0.5, 0)
(0.6, 0)
(0.7, 0)
(0.8, 0)
(0.9, 0)
(1, 0)};
\end{axis}
\end{tikzpicture}
    \caption{Histogram of Kendall $\tau_C$ values for the 100 assessors}
    \label{Fig2}
\end{figure}

Figure~\ref{Fig2} presents the distribution of Kendall $\tau_C$ coefficients for all assessors. Unsurprisingly, they are typically negative, but some positive values also occur. At a significance level 5\%, 88 assessors have a significantly negative Kendall $\tau_C$ coefficient, the remaining 12 assessors get the ``inconsistent'' label.

\subsection{The connection between consistency and the use of the scales}

\begin{figure}[t!]
\center
\begin{subfigure}{0.5\textwidth}
\caption{Liking scores}
\label{Fig3a}
\begin{tikzpicture}

\draw[very thick](-0.5,0) -- (5.5,0);

\draw[->, very thick](0,-0.5) -- (0,5.5);

\draw[very thick](-0.1,1) -- (0.1,1);
\node at (-0.3,1) {3};
\draw[very thick](-0.1,2) -- (0.1,2);
\node at (-0.3,2) {4};
\draw[very thick](-0.1,3) -- (0.1,3);
\node at (-0.3,3) {5};
\draw[very thick](-0.1,4) -- (0.1,4);
\node at (-0.3,4) {6};
\draw[very thick](-0.1,5) -- (0.1,5);
\node at (-0.3,5) {7};

\filldraw[fill=blue!25]
(1,3.4) -- (1,4.06) -- (2,4.06) --  (2,3.4) -- (1,3.4);

\draw[very thick](1,3.72) -- (2,3.72);

\draw[dashed] (1.5,3.3)--(1.5,3.4);
\draw (1.2,3.3)--(1.8,3.3);

\draw[dashed] (1.5,4.06)--(1.5,4.9);
\draw (1.2,4.9)--(1.8,4.9);

\draw (1.45,2.28)--(1.55,2.38);
\draw (1.55,2.28)--(1.45,2.38);

\draw (1.5,0)--(1.5,-0.2);
\node at (1.5,-0.5) {\scriptsize{Inconsistent}};

\filldraw[fill=blue!25]
(3,3.28) -- (3,4.01) -- (4,4.01) --  (4,3.28) -- (3,3.28);

\draw[very thick](3,3.61) -- (4,3.61);

\draw[dashed] (3.5,2.29)--(3.5,3.28);
\draw (3.2,2.29)--(3.8,2.29);

\draw[dashed] (3.5,4.01)--(3.5,5.0);
\draw (3.2,5.0)--(3.8,5.0);

\draw (3.5,0)--(3.5,-0.2);
\node at (3.5,-0.5) {\scriptsize{Consistent}};
\end{tikzpicture}
\end{subfigure}

\begin{subfigure}{0.5\textwidth}
\caption{JAR scores}
\label{Fig3b}
\begin{tikzpicture}
\draw[very thick](7.5,0) -- (13.5,0);

\draw[->, very thick](8,-0.5) -- (8,5.5);

\draw[very thick](7.9,0.3) -- (8.1,0.3);
\node at (7.5,0.3) {-0.5};
\draw[very thick](7.9,1.55) -- (8.1,1.55);
\node at (7.5,1.55) {-0.25};
\draw[very thick](7.9,2.8) -- (8.1,2.8);
\node at (7.5,2.8) {0};
\draw[very thick](7.9,4.05) -- (8.1,4.05);
\node at (7.5,4.05) {0.25};
\draw[very thick](7.9,5.3) -- (8.1,5.3);
\node at (7.5,5.3) {0.5};

\filldraw[fill=blue!25]
(9,2.44) -- (9,3.16) -- (10,3.16) --  (10,2.44) -- (9,2.44);

\draw[very thick](9,2.72) -- (10,2.72);

\draw[dashed] (9.5,2.0)--(9.5,2.44);
\draw (9.2,2.0)--(9.8,2.0);

\draw[dashed] (9.5,3.17)--(9.5,3.33);
\draw (9.2,3.33)--(9.8,3.33);

\draw (9.45,0.89)--(9.55,0.99);
\draw (9.55,0.89)--(9.45,0.99);

\draw (9.45,4.39)--(9.55,4.49);
\draw (9.55,4.39)--(9.45,4.49);

\draw (9.45,4.73)--(9.55,4.83);
\draw (9.55,4.73)--(9.45,4.83);

\draw (9.5,0)--(9.5,-0.2);
\node at (9.5,-0.5) {\scriptsize{Inconsistent}};

\filldraw[fill=blue!25]
(11,2.22) -- (11,3.58) -- (12,3.58) --  (12,2.22) -- (11,2.22);

\draw[very thick](11,2.91) -- (12,2.91);

\draw[dashed] (11.5,0.63)--(11.5,2.22);
\draw (11.2,0.63)--(11.8,0.63);

\draw[dashed] (11.5,3.58)--(11.5,5.30);
\draw (11.2,5.3)--(11.8,5.3);

\draw (11.45,0.02)--(11.55,0.12);
\draw (11.55,0.02)--(11.45,0.12);

\draw (11.5,0)--(11.5,-0.2);
\node at (11.5,-0.5) {\scriptsize{Consistent}};

\end{tikzpicture}
\end{subfigure}
\caption{Comparison of mean scores for the two types of assessors}
\label{Fig3}
\end{figure}
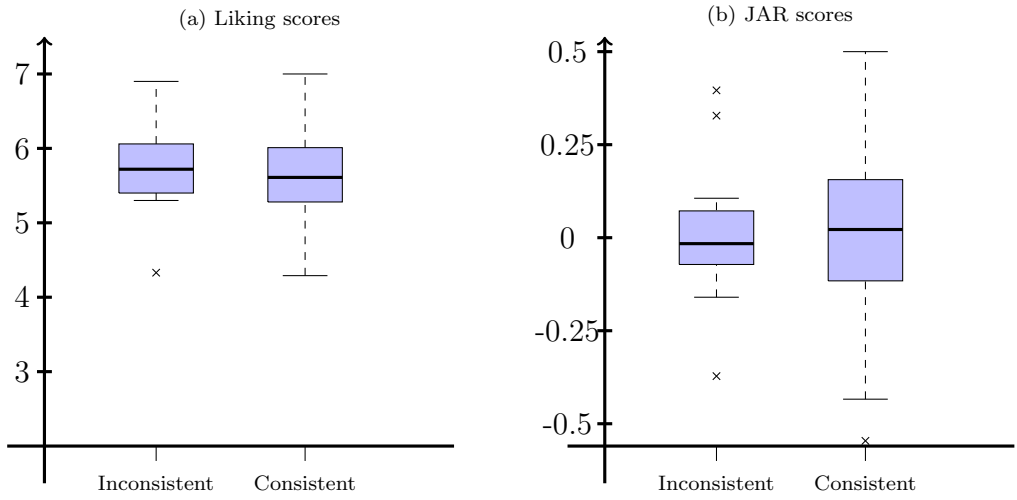 

In the first step, the mean liking score and mean JAR scores have been calculated across all samples and attributes that can be seen in Figure~\ref{Fig3}. Interestingly, three is no significant difference between the two groups, statistical tests do not reject a null hypothesis of equal means for consistent and inconsistent assessors.

\begin{figure}[t!]
\center
\begin{subfigure}{0.5\textwidth}
\caption{Liking scores}
\label{Fig4a}
\begin{tikzpicture}

\draw[very thick](-0.5,0) -- (5.5,0);

\draw[->, very thick](0,-0.5) -- (0,5.5);

\draw[very thick](-0.1,1) -- (0.1,1);
\node at (-0.4,1) {1.4};
\draw[very thick](-0.1,2) -- (0.1,2);
\node at (-0.4,2) {1.8};
\draw[very thick](-0.1,3) -- (0.1,3);
\node at (-0.4,3) {2.2};
\draw[very thick](-0.1,4) -- (0.1,4);
\node at (-0.4,4) {2.6};
\draw[very thick](-0.1,5) -- (0.1,5);
\node at (-0.4,5) {3};

\filldraw[fill=blue!25]
(1,1.17) -- (1,2.51) -- (2,2.51) --  (2,1.17) -- (1,1.17);

\draw[very thick](1,1.68) -- (2,1.68);

\draw[dashed] (1.5,0.92)--(1.5,1.17);
\draw (1.2,0.92)--(1.8,0.92);

\draw[dashed] (1.5,2.51)--(1.5,3.4);
\draw (1.2,3.4)--(1.8,3.4);

\draw (1.5,0)--(1.5,-0.2);
\node at (1.5,-0.5) {\scriptsize{Inconsistent}};

\filldraw[fill=blue!25]
(3,1.99) -- (3,2.86) -- (4,2.86) --  (4,1.99) -- (3,1.99);

\draw[very thick](3,2.42) -- (4,2.42);

\draw[dashed] (3.5,0.86)--(3.5,1.99);
\draw (3.2,0.86)--(3.8,0.86);

\draw[dashed] (3.5,2.86)--(3.5,4.12);
\draw (3.2,4.12)--(3.8,4.12);

\draw (3.45,0.52)--(3.55,0.62);
\draw (3.55,0.52)--(3.45,0.62);

\draw (3.45,4.26)--(3.55,4.36);
\draw (3.55,4.26)--(3.45,4.36);

\draw (3.45,4.54)--(3.55,4.64);
\draw (3.55,4.54)--(3.45,4.64);

\draw (3.5,0)--(3.5,-0.2);
\node at (3.5,-0.5) {\scriptsize{Consistent}};
\end{tikzpicture}
\end{subfigure}

\begin{subfigure}{0.5\textwidth}
\caption{JAR scores}
\label{Fig4b}

\begin{tikzpicture}
\draw[very thick](7.5,0) -- (13.5,0);

\draw[->, very thick](8,-0.5) -- (8,5.5);

\draw[very thick](7.9,0.3) -- (8.1,0.3);
\node at (7.6,0.3) {0.5};
\draw[very thick](7.9,1.55) -- (8.1,1.55);
\node at (7.6,1.55) {0.7};
\draw[very thick](7.9,2.8) -- (8.1,2.8);
\node at (7.6,2.8) {0.9};
\draw[very thick](7.9,4.05) -- (8.1,4.05);
\node at (7.6,4.05) {1.1};
\draw[very thick](7.9,5.3) -- (8.1,5.3);
\node at (7.6,5.3) {1.3};

\filldraw[fill=blue!25]
(9,1.97) -- (9,2.96) -- (10,2.96) --  (10,1.97) -- (9,1.97);

\draw[very thick](9,2.54) -- (10,2.54);

\draw[dashed] (9.5,1.68)--(9.5,1.97);
\draw (9.2,1.68)--(9.8,1.68);

\draw[dashed] (9.5,2.96)--(9.5,3.91);
\draw (9.2,3.91)--(9.8,3.91);

\draw (9.5,0)--(9.5,-0.2);
\node at (9.5,-0.5) {\scriptsize{Inconsistent}};

\filldraw[fill=blue!25]
(11,3.00) -- (11,4.06) -- (12,4.06) --  (12,3.00) -- (11,3.00);

\draw[very thick](11,3.49) -- (12,3.49);

\draw[dashed] (11.5,1.57)--(11.5,3.00);
\draw (11.2,1.57)--(11.8,1.57);

\draw[dashed] (11.5,4.06)--(11.5,5.02);
\draw (11.2,5.02)--(11.8,5.02);

\draw (11.45,0.5)--(11.55,0.6);
\draw (11.55,0.5)--(11.45,0.6);

\draw (11.45,1.2)--(11.55,1.3);
\draw (11.55,1.2)--(11.45,1.3);

\draw (11.45,1.25)--(11.55,1.35);
\draw (11.55,1.25)--(11.45,1.35);

\draw (11.5,0)--(11.5,-0.2);
\node at (11.5,-0.5) {\scriptsize{Consistent}};

\end{tikzpicture}
\end{subfigure}
\caption{Comparison of standard deviations for the two types of assessors}
\label{Fig4}
\end{figure}
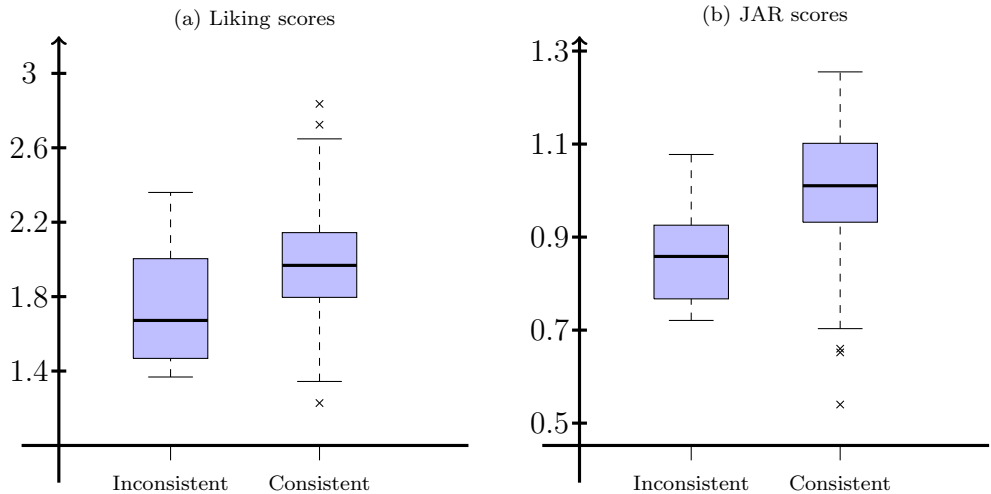
  
In the second step, the standard deviation of liking scores and JAR scores have been calculated across all samples and attributes that can be seen in Figure~\ref{Fig4}. Now, there are significant differences between the two groups, the consistent assessors generally use both scales with a higher spread than the inconsistent assessors. 
 
This finding is somewhat surprising because we assumed that inconsistent assessors would only select a few random values when the standard deviations of the inconsistent assessors would be higher. Looking at the results, it seems that the consistent assessors are more confident and, therefore, use the scale more ``intensively'', giving values from a broader range of the scale.

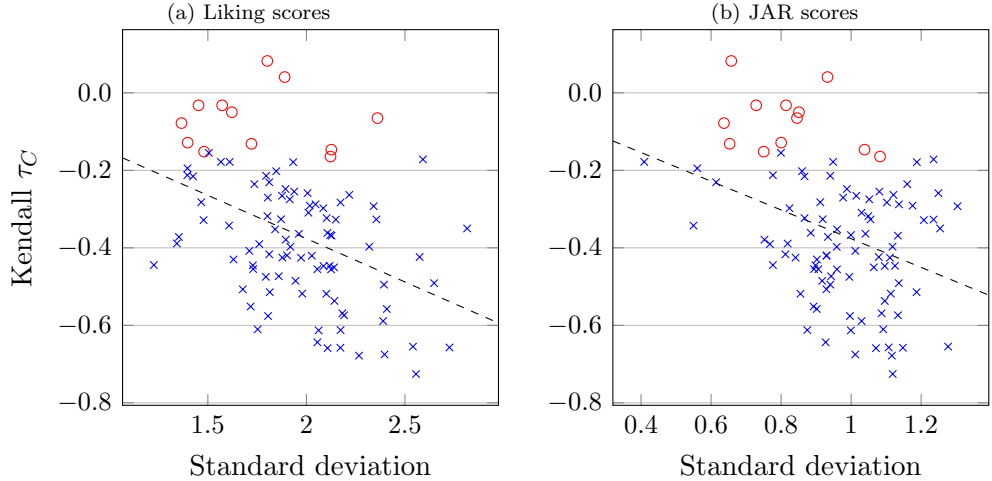
\begin{figure}[t!]
\centering

\begin{subfigure}{0.48\textwidth}
\caption{Liking scores}
\label{Fig5a}
\begin{tikzpicture}
\begin{axis}[
xlabel = Standard deviation,
x label style = {font=\small},
x tick label style = {font=\footnotesize},
ylabel = Kendall $\tau_C$,
y label style = {font=\small},
y tick label style = {font=\footnotesize,/pgf/number format/.cd,fixed,fixed zerofill,precision=1},
width = \textwidth,
height = \textwidth,
ymajorgrids = true,
] 
\addplot [red, mark=o, only marks] coordinates {
(1.88961945187619,0.0407407407407407)
(2.36072736838991,-0.0651851851851852)
(1.80168460495458,0.0822222222222222)
(1.5725882194221,-0.0322222222222222)
(2.12237941519005,-0.164444444444444)
(2.12649301299865,-0.146666666666667)
(1.36612303063504,-0.0781481481481481)
(1.39751875717437,-0.128518518518519)
(1.72116878095506,-0.131481481481481)
(1.45257960732123,-0.0322222222222222)
(1.62183139649339,-0.05)
(1.48079676228239,-0.151851851851852)
};
\addplot [blue, mark=x, only marks] coordinates {
(2.08702668873702,-0.447037037037037)
(1.91872955262847,-0.397407407407407)
(2.12790155874171,-0.368518518518519)
(1.84712213315241,-0.202222222222222)
(1.93286188233401,-0.178888888888889)
(2.18255482589555,-0.568888888888889)
(2.12084947529967,-0.367037037037037)
(1.46643682230355,-0.282222222222222)
(1.84143595924756,-0.352222222222222)
(2.10296740586067,-0.323333333333333)
(2.00589020278229,-0.258888888888889)
(2.06196154019866,-0.612962962962963)
(2.17338342351689,-0.612222222222222)
(1.22606930577834,-0.444444444444444)
(2.35514153193724,-0.326666666666667)
(1.85975177219495,-0.473333333333333)
(2.14926750579951,-0.327037037037037)
(2.72584156449098,-0.657037037037037)
(1.79404619444852,-0.474814814814815)
(2.54068025925642,-0.654814814814815)
(2.17318236648526,-0.282962962962963)
(2.11775680441516,-0.446296296296296)
(2.31900361745681,-0.397037037037037)
(1.81411444951939,-0.514444444444444)
(2.14214006193428,-0.537037037037037)
(1.94480645818556,-0.485185185185185)
(2.05556399093994,-0.455185185185185)
(1.90196691672321,-0.41962962962963)
(2.38934902516249,-0.588888888888889)
(2.10643743122966,-0.361481481481481)
(2.1724928868286,-0.657777777777778)
(2.26698059073111,-0.678148148148148)
(2.10109655693712,-0.518518518518518)
(2.81568639724676,-0.35)
(2.01706205304713,-0.291111111111111)
(2.4080856306399,-0.557777777777778)
(1.96168034164502,-0.363703703703704)
(1.75265092600166,-0.61037037037037)
(2.10711890087436,-0.658888888888889)
(2.12517165349155,-0.455555555555556)
(1.72822641509228,-0.445185185185185)
(1.39662514575415,-0.194814814814815)
(1.91531054450646,-0.274814814814815)
(1.87619292055909,-0.265555555555556)
(2.39370794017804,-0.494814814814815)
(2.55623396521013,-0.725555555555556)
(2.34110530144618,-0.292592592592593)
(2.02731041422015,-0.42037037037037)
(2.21739800951121,-0.263333333333333)
(2.19137453372676,-0.574074074074074)
(1.87778922792718,-0.425185185185185)
(1.73010358850743,-0.454814814814815)
(2.01024218887367,-0.309259259259259)
(1.60756637900939,-0.342592592592593)
(2.04643351646916,-0.287777777777778)
(1.80248129711987,-0.318148148148148)
(2.14053676164766,-0.45037037037037)
(1.87129575835328,-0.326296296296296)
(2.39652265771447,-0.675185185185185)
(1.73601060882633,-0.235555555555556)
(1.7957155279104,-0.214074074074074)
(2.64808601645826,-0.491111111111111)
(1.8054225438378,-0.575925925925926)
(1.34247800392082,-0.388888888888889)
(1.61070854671847,-0.178148148148148)
(1.67708355142251,-0.507037037037037)
(1.97310252724516,-0.425555555555556)
(2.57462892614951,-0.423703703703704)
(1.89407381204987,-0.248148148148148)
(2.59125582325888,-0.171851851851852)
(1.47830756557185,-0.328518518518519)
(1.81118731792828,-0.416666666666667)
(1.81149747339869,-0.230740740740741)
(1.62997055034937,-0.43)
(1.71050948451893,-0.407777777777778)
(1.8040390250754,-0.27037037037037)
(2.05568545663708,-0.643703703703704)
(1.93837647808327,-0.254814814814815)
(1.71706569668769,-0.551481481481481)
(2.08568032593025,-0.297777777777778)
(1.56622433790763,-0.178148148148148)
(1.76085894373324,-0.39037037037037)
(1.97878889554535,-0.518148148148148)
(1.89565506220315,-0.378888888888889)
(1.35285336777739,-0.371851851851852)
(1.39577568498542,-0.212592592592593)
(1.50488177726543,-0.154814814814815)
(1.42459228238686,-0.215555555555556)
};
\draw [black,dashed] (\pgfkeysvalueof{/pgfplots/xmin},0.07062-0.22345*\pgfkeysvalueof{/pgfplots/xmin}) -- (\pgfkeysvalueof{/pgfplots/xmax},0.07062-0.22345*\pgfkeysvalueof{/pgfplots/xmax});
\end{axis}
\end{tikzpicture}
\end{subfigure}
\hspace{0.02\textwidth}
\begin{subfigure}{0.48\textwidth}
\caption{JAR scores}
\label{Fig5b}
\begin{tikzpicture}
\begin{axis}[
xlabel = Standard deviation,
x label style = {font=\small},
x tick label style = {font=\footnotesize},
y tick label style = {font=\footnotesize,/pgf/number format/.cd,fixed,fixed zerofill,precision=1},
width = \textwidth,
height = \textwidth,
ymajorgrids = true,
] 
\addplot [red, mark=o, only marks] coordinates {
(0.932490255929383,0.0407407407407407)
(0.845344155698159,-0.0651851851851852)
(0.657711387816055,0.0822222222222222)
(0.728927863517533,-0.0322222222222222)
(1.08266329033993,-0.164444444444444)
(1.03834113354571,-0.146666666666667)
(0.636881446996291,-0.0781481481481481)
(0.800202845694252,-0.128518518518519)
(0.653044344860712,-0.131481481481481)
(0.814123150616378,-0.0322222222222222)
(0.851011001893231,-0.05)
(0.750280846293311,-0.151851851851852)
};
\addplot [blue, mark=x, only marks] coordinates {
(1.09595784385423,-0.447037037037037)
(0.958827832035367,-0.397407407407407)
(1.13375185046242,-0.368518518518519)
(0.859767991234589,-0.202222222222222)
(1.18774432403439,-0.178888888888889)
(1.08697888485924,-0.568888888888889)
(0.999438044350115,-0.367037037037037)
(0.911844654619033,-0.282222222222222)
(0.959868904906676,-0.352222222222222)
(0.866925919161123,-0.323333333333333)
(1.24921948165977,-0.258888888888889)
(0.999375585327815,-0.612962962962963)
(0.874739126440473,-0.612222222222222)
(0.776447944640579,-0.444444444444444)
(1.05551284492078,-0.326666666666667)
(0.942875247757601,-0.473333333333333)
(1.23535111790914,-0.327037037037037)
(1.1072905601148,-0.657037037037037)
(0.994930972705865,-0.474814814814815)
(1.27709095148976,-0.654814814814815)
(1.10209198417057,-0.282962962962963)
(1.12540740299751,-0.446296296296296)
(1.11822938335913,-0.397037037037037)
(1.18695573636925,-0.514444444444444)
(1.09595784385423,-0.537037037037037)
(0.917372897848996,-0.485185185185185)
(0.906420433627177,-0.455185185185185)
(0.930077252372497,-0.41962962962963)
(1.02989036178778,-0.588888888888889)
(0.884673211776787,-0.361481481481481)
(1.14797764967129,-0.657777777777778)
(1.11627389816762,-0.678148148148148)
(0.855400702622719,-0.518518518518518)
(1.25420640560682,-0.35)
(1.17554141907678,-0.291111111111111)
(0.901933005309937,-0.557777777777778)
(1.04104288439097,-0.363703703703704)
(1.09184927140181,-0.61037037037037)
(1.07083688555059,-0.658888888888889)
(0.959608742575049,-0.455555555555556)
(0.898326923187681,-0.445185185185185)
(0.56045282209367,-0.194814814814815)
(1.05178048910095,-0.274814814814815)
(1.01480919530294,-0.265555555555556)
(0.940488889254671,-0.494814814814815)
(1.11873166993402,-0.725555555555556)
(1.30379260469406,-0.292592592592593)
(0.929540178701397,-0.42037037037037)
(1.12090564418734,-0.263333333333333)
(1.13375185046242,-0.574074074074074)
(0.841273038076377,-0.425185185185185)
(0.893938528302992,-0.454814814814815)
(1.02989036178778,-0.309259259259259)
(0.549770125608182,-0.342592592592593)
(1.13666617241634,-0.287777777777778)
(1.04910639104208,-0.318148148148148)
(1.06475718601404,-0.45037037037037)
(0.918732775408192,-0.326296296296296)
(1.01185235756414,-0.675185185185185)
(1.16030905521782,-0.235555555555556)
(0.940488889254671,-0.214074074074074)
(1.13622675375783,-0.491111111111111)
(0.996623513151427,-0.575925925925926)
(0.818710657725706,-0.388888888888889)
(0.94842006738561,-0.178148148148148)
(0.929473022672071,-0.507037037037037)
(1.11083642047723,-0.425555555555556)
(1.07890914530382,-0.423703703703704)
(0.988131820346872,-0.248148148148148)
(1.23535111790914,-0.171851851851852)
(1.20775686882016,-0.328518518518519)
(0.812280893341638,-0.416666666666667)
(0.614433145173371,-0.230740740740741)
(0.902278984457002,-0.43)
(1.01203741301561,-0.407777777777778)
(0.97720588200758,-0.27037037037037)
(0.927119497043263,-0.643703703703704)
(1.08075896945731,-0.254814814814815)
(0.892540874964921,-0.551481481481481)
(0.823803183650839,-0.297777777777778)
(0.409012085875234,-0.178148148148148)
(0.767960152940483,-0.39037037037037)
(1.11336226385969,-0.518148148148148)
(0.75252384336409,-0.378888888888889)
(0.933761271610442,-0.371851851851852)
(0.776447944640579,-0.212592592592593)
(0.799578540667913,-0.154814814814815)
(0.866997919985064,-0.215555555555556)
};
\draw [black,dashed] (\pgfkeysvalueof{/pgfplots/xmin},-0.005823-0.370524*\pgfkeysvalueof{/pgfplots/xmin}) -- (\pgfkeysvalueof{/pgfplots/xmax},-0.005823-0.370524*\pgfkeysvalueof{/pgfplots/xmax});
\end{axis}
\end{tikzpicture}
\end{subfigure}

\caption{The relationship between the standard deviation of scores and Kendall $\tau_C$ coefficients for the assessors \\
\vspace{0.2cm}
\scriptsize{\emph{Notes:} Blue points represent consistent assessors. Red points represent inconsistent assessors. \\
The dashed line shows the linear trend, which is $y=0.07062-0.22345x$ ($R^2=0.1825$) for Figure~\ref{Fig5a} and $y=-0.005823-0.370524x$ ($R^2=0.1318$) for Figure~\ref{Fig5b}.}}
\label{Fig5}

\end{figure}


The relationship between the standard deviation of the liking/JAR scores and Kendall $\tau_C$ coefficient values is worth further analyzing.
Figure~\ref{Fig5} reinforces that for higher standard deviation the Kendall $\tau_C$ correlation coefficient tends to be lower for a higher standard deviation, which makes the assessor is less inconsistent. The slope of the fitted curve is also negative.

\begin{figure}[t!]
\centering

\begin{tikzpicture}
\begin{axis}[
xlabel = Number of concordant and disconcordant evaluation pairs,
x label style = {font=\small},
x tick label style = {font=\footnotesize},
ylabel = Kendall $\tau_C$,
y label style = {font=\small},
y tick label style = {font=\footnotesize,/pgf/number format/.cd,fixed,fixed zerofill,precision=1},
width = 0.8\textwidth,
height = 0.6\textwidth,
ymajorgrids = true,
] 
\addplot [red, mark=o, only marks] coordinates {
(2144,0.0407407407407407)
(1890,-0.0651851851851852)
(1428,0.0822222222222222)
(1749,-0.0322222222222222)
(2290,-0.164444444444444)
(2180,-0.146666666666667)
(1585,-0.0781481481481481)
(1509,-0.128518518518519)
(1555,-0.131481481481481)
(1723,-0.0322222222222222)
(1779,-0.05)
(1740,-0.151851851851852)
};
\addplot [blue, mark=x, only marks] coordinates {
(2245,-0.447037037037037)
(2045,-0.397407407407407)
(2265,-0.368518518518519)
(2008,-0.202222222222222)
(2149,-0.178888888888889)
(2276,-0.568888888888889)
(2129,-0.367037037037037)
(1948,-0.282222222222222)
(2115,-0.352222222222222)
(1921,-0.323333333333333)
(2285,-0.258888888888889)
(2209,-0.612962962962963)
(2193,-0.612222222222222)
(1788,-0.444444444444444)
(2270,-0.326666666666667)
(2118,-0.473333333333333)
(2393,-0.327037037037037)
(2348,-0.657037037037037)
(2060,-0.474814814814815)
(2488,-0.654814814814815)
(2248,-0.282962962962963)
(2271,-0.446296296296296)
(2324,-0.397037037037037)
(2299,-0.514444444444444)
(2240,-0.537037037037037)
(2166,-0.485185185185185)
(2127,-0.455185185185185)
(1861,-0.41962962962963)
(2232,-0.588888888888889)
(2082,-0.361481481481481)
(2392,-0.657777777777778)
(2455,-0.678148148148148)
(2116,-0.518518518518518)
(2363,-0.35)
(2248,-0.291111111111111)
(2154,-0.557777777777778)
(2196,-0.363703703703704)
(2154,-0.61037037037037)
(2343,-0.658888888888889)
(2092,-0.455555555555556)
(2050,-0.445185185185185)
(1244,-0.194814814814815)
(2222,-0.274814814814815)
(2125,-0.265555555555556)
(2018,-0.494814814814815)
(2365,-0.725555555555556)
(2378,-0.292592592592593)
(2045,-0.42037037037037)
(2251,-0.263333333333333)
(2208,-0.574074074074074)
(1894,-0.425185185185185)
(2010,-0.454814814814815)
(2117,-0.309259259259259)
(1583,-0.342592592592593)
(2229,-0.287777777777778)
(2041,-0.318148148148148)
(2158,-0.45037037037037)
(2017,-0.326296296296296)
(2319,-0.675185185185185)
(2048,-0.235555555555556)
(1992,-0.214074074074074)
(2308,-0.491111111111111)
(2189,-0.575925925925926)
(1714,-0.388888888888889)
(1999,-0.178148148148148)
(1881,-0.507037037037037)
(2217,-0.425555555555556)
(2274,-0.423703703703704)
(2128,-0.248148148148148)
(2386,-0.171851851851852)
(2153,-0.328518518518519)
(1877,-0.416666666666667)
(1537,-0.230740740740741)
(1993,-0.43)
(2145,-0.407777777777778)
(2150,-0.27037037037037)
(2264,-0.643703703703704)
(2248,-0.254814814814815)
(2021,-0.551481481481481)
(1858,-0.297777777777778)
(911,-0.178148148148148)
(1966,-0.39037037037037)
(2319,-0.518148148148148)
(1779,-0.378888888888889)
(1956,-0.371851851851852)
(1656,-0.212592592592593)
(1742,-0.154814814814815)
(1840,-0.215555555555556)
};
\draw [black,dashed] (\pgfkeysvalueof{/pgfplots/xmin},0.3459+-0.0003*\pgfkeysvalueof{/pgfplots/xmin}) -- (\pgfkeysvalueof{/pgfplots/xmax},0.3459+-0.0003*\pgfkeysvalueof{/pgfplots/xmax});
\end{axis}
\end{tikzpicture}

\caption{The relationship between the number of concordant and disconcordant evaluation pairs and Kendall $\tau_C$ coefficients for the assessors \\
\vspace{0.2cm}
\scriptsize{\emph{Notes:} Blue points represent consistent assessors. Red points represent inconsistent assessors. \\
The dashed line shows the linear trend, which is $y=0.3459-0.0003x$ ($R^2=0.2861$).}}
\label{Fig6}

\end{figure}


Finally, Figure~\ref{Fig6} compares the number of evaluation pairs without ties, and the Kendall $\tau_C$ coefficients. The inconsistent assessors seem to be less discriminative in their evaluations, and there is non-negligible connection between the number of evaluation pairs without ties and Kendall $\tau_C$ even for the consistent assessors. This provides another evidence that the less inconsistent assessors use the two scales better, they are more likely to avoid giving the same score for two different products.

\subsection{The consistency of evaluations with respect to attributes}

Any attribute is only one aspect of the sensory evaluation. As such, it does not include a direct source of inconsistency. However, once the assessments are collected from two different scales (nine-category monotonic ascending liking response scale, five-category just about right (JAR) scale of intensity), it becomes reasonable to define the level of inconsistency with respect to a particular attribute. If the same biscuit is evaluated differently by the two scales (e.g.\ it gets the highest liking score and the highest/lowest JAR score), then the assessment, with respect to the given attribute, becomes controversial.

\begin{table}[t!]
    \centering
    \caption{Kendall $\tau_C$ correlation coefficients for each attribute}
    \label{Table5}
    \rowcolors{1}{}{gray!20}
    \begin{tabularx}{0.8\textwidth}{lcC} \toprule
    Attribute & Abbreviation & Value of Kendall $\tau_C$ \\ \bottomrule
    Colour & CL & $-$0.54 \\ 
    Global odour& GO & $-$0.35 \\ 
    Sweet odour & SO & $-$0.35 \\ 
    Margarine odour& MO & $-$0.20 \\ 
    Global taste& GT & $-$0.47 \\ 
    Sweet taste& ST & $-$0.39 \\ 
    Margarine taste& MT & $-$0.29 \\ 
    Hardness & HN & $-$0.43 \\ 
    Crumbleness& CR & $-$0.31 \\ \toprule 
    \end{tabularx}
\end{table}

\begin{figure}[t!]
\centering

\begin{subfigure}{0.47\textwidth}
\caption{Liking scores}
\label{Fig7a}
\begin{tikzpicture}
\begin{axis}[
xlabel = Standard deviation,
x label style = {font=\small},
x tick label style = {font=\footnotesize},
ylabel = Kendall $\tau_C$,
y label style = {font=\small},
y tick label style = {font=\footnotesize,/pgf/number format/.cd,fixed,fixed zerofill,precision=1},
width = \textwidth,
height = \textwidth,
ymajorgrids = true,
] 
\addplot [blue, mark=x, only marks] coordinates {
(1.9872022979488,-0.536412)
(1.92790477379377,-0.353391)
(1.92243979743834,-0.35157)
(1.6323260974721,-0.200721)
(2.15859642789905,-0.472104)
(1.9651699399525,-0.391551)
(1.85008075466959,-0.290193)
(2.45088674684202,-0.426651)
(2.11736492703413,-0.309345)
};
\draw [black,dashed] (\pgfkeysvalueof{/pgfplots/xmin},0.07062-0.22345*\pgfkeysvalueof{/pgfplots/xmin}) -- (\pgfkeysvalueof{/pgfplots/xmax},0.07062-0.22345*\pgfkeysvalueof{/pgfplots/xmax});
\node[pin={[pin distance=0.2cm, ultra thick, pin edge={blue}] 90:{\textcolor{blue}{\scriptsize{CL}}}}] at (1.9872022979488,-0.536412) {};
\node[pin={[pin distance=0.2cm, ultra thick, pin edge={blue}] 90:{\textcolor{blue}{\scriptsize{GO}}}}] at (1.92790477379377,-0.353391) {};
\node[pin={[pin distance=0.2cm, ultra thick, pin edge={blue}] 0:{\textcolor{blue}{\scriptsize{SO}}}}] at (1.92243979743834,-0.35157) {};
\node[pin={[pin distance=0.2cm, ultra thick, pin edge={blue}] 270:{\textcolor{blue}{\scriptsize{MO}}}}] at (1.6323260974721,-0.200721) {};
\node[pin={[pin distance=0.2cm, ultra thick, pin edge={blue}] 270:{\textcolor{blue}{\scriptsize{GT}}}}] at (2.15859642789905,-0.472104) {};
\node[pin={[pin distance=0.2cm, ultra thick, pin edge={blue}] 270:{\textcolor{blue}{\scriptsize{ST}}}}] at (1.9651699399525,-0.391551) {};
\node[pin={[pin distance=0.2cm, ultra thick, pin edge={blue}] 90:{\textcolor{blue}{\scriptsize{MT}}}}] at (1.85008075466959,-0.290193) {};
\node[pin={[pin distance=0.2cm, ultra thick, pin edge={blue}] 90:{\textcolor{blue}{\scriptsize{HN}}}}] at (2.45088674684202,-0.426651) {};
\node[pin={[pin distance=0.2cm, ultra thick, pin edge={blue}] 90:{\textcolor{blue}{\scriptsize{CR}}}}] at (2.11736492703413,-0.309345) {};
\end{axis}
\end{tikzpicture}
\end{subfigure}
\hspace{0.02\textwidth}
\begin{subfigure}{0.47\textwidth}
\caption{JAR scores}
\label{Fig7b}
\begin{tikzpicture}
\begin{axis}[
xlabel = Standard deviation,
x label style = {font=\small},
x tick label style = {font=\footnotesize},
y tick label style = {font=\footnotesize,/pgf/number format/.cd,fixed,fixed zerofill,precision=1},
width = \textwidth,
height = \textwidth,
ymajorgrids = true,
] 
\addplot [blue, mark=x, only marks] coordinates {
(1.0286311691185,-0.536412)
(1.02519023357376,-0.353391)
(0.941359207293984,-0.35157)
(0.941065674968865,-0.200721)
(0.927631122537132,-0.472104)
(0.922519399276808,-0.391551)
(0.890200233911706,-0.290193)
(0.944376962789673,-0.426651)
(0.862686633852773,-0.309345)
};
\draw [black,dashed] (\pgfkeysvalueof{/pgfplots/xmin},0.4312-0.8502*\pgfkeysvalueof{/pgfplots/xmin}) -- (\pgfkeysvalueof{/pgfplots/xmax},0.4312-0.8502*\pgfkeysvalueof{/pgfplots/xmax});
\node[pin={[pin distance=0.2cm, ultra thick, pin edge={blue}] 90:{\textcolor{blue}{\scriptsize{CL}}}}] at (1.0286311691185,-0.536412) {};
\node[pin={[pin distance=0.2cm, ultra thick, pin edge={blue}] 90:{\textcolor{blue}{\scriptsize{GO}}}}] at (1.02519023357376,-0.353391) {};
\node[pin={[pin distance=0.2cm, ultra thick, pin edge={blue}] 90:{\textcolor{blue}{\scriptsize{SO}}}}] at (0.941359207293984,-0.35157) {};
\node[pin={[pin distance=0.2cm, ultra thick, pin edge={blue}] 270:{\textcolor{blue}{\scriptsize{MO}}}}] at (0.941065674968865,-0.200721) {};
\node[pin={[pin distance=0.2cm, ultra thick, pin edge={blue}] 270:{\textcolor{blue}{\scriptsize{GT}}}}] at (0.927631122537132,-0.472104) {};
\node[pin={[pin distance=0.2cm, ultra thick, pin edge={blue}] 270:{\textcolor{blue}{\scriptsize{ST}}}}] at (0.922519399276808,-0.391551) {};
\node[pin={[pin distance=0.2cm, ultra thick, pin edge={blue}] 90:{\textcolor{blue}{\scriptsize{MT}}}}] at (0.890200233911706,-0.290193) {};
\node[pin={[pin distance=0.2cm, ultra thick, pin edge={blue}] 270:{\textcolor{blue}{\scriptsize{HN}}}}] at (0.944376962789673,-0.426651) {};
\node[pin={[pin distance=0.2cm, ultra thick, pin edge={blue}] 90:{\textcolor{blue}{\scriptsize{CR}}}}] at (0.862686633852773,-0.309345) {};
\end{axis}
\end{tikzpicture}
\end{subfigure}

\caption{The relationship between the standard deviation of scores and Kendall $\tau_C$ coefficients for the attributes \\ \vspace{0.2cm}
\scriptsize{\emph{Notes:} Each point represents an attribute. \\
The dashed line shows the linear trend, which is $y=0.1566-0.2632x$ ($R^2=0.3523$) for Figure~\ref{Fig7a} and $y=0.4312-0.8502x$ ($R^2=0.2138$) Figure~\ref{Fig7b}. \\
The abbreviations of the attributes are given in table~\ref{Table5}.}}
\label{Fig7}

\end{figure}


The Kendall $\tau_C$ coefficients can also be calculated between the rankings of liking scores and JAR scores (in absolute value) for each attribute across all assessors and samples, which is presented in Table~\ref{Table5}.
Furthermore, $\tau_C$ is plotted as a function of the standard deviations of the scores in Figure~\ref{Fig7}.
The standard deviations are the smallest and the Kendall $\tau_C$ coefficients are the highest for margarine odour and margarine taste that the assessors probably find difficult to evaluate. On the other hand, colour and global taste are evaluated less inconsistently as shown by the low value of Kendall $\tau_C$ (Table~\ref{Table5}). Again, Figure~\ref{Fig7} reveals a (significant) negative correlation between the standard deviation of scores and Kendall $\tau_C$ coefficients, although the sample size is small, making the result less reliable.

\subsection{The effect of inconsistent evaluations}

\begin{figure}[t!]
\center
\includestandalone[width=\textwidth]{figure8}
\caption{The relationship between global taste and global liking scores \\ \vspace{0.2cm}
\scriptsize{\emph{Notes:} The ratio of consistent and inconsistent assessors is shown by blue and orange circular sectors.
The three lines shows linear trends, which are $y=1.21097+0.76649x$ ($R^2=0.5619$) for consistent assessors (dashed blue); $y=2.53432+0.58105x$ ($R^2=0.3428$) for inconsistent assessors (solid red); $y=1.34029+0.74841x$ ($R^2=0.5381$) for all assessors (black dotted).}}
\label{Fig8}
\end{figure}

Here we focus on the importance of considering the consistency of the assessors. The relationship between global taste and global liking scores can be studied by fitting three models. In Figure~\ref{Fig8}, the first regression is based on all evaluations (black dotted line), the second only on consistent assessors (blue dashed line), and the third only on inconsistent assessors (red solid line). The slope is somewhat steeper for consistent assessors but the blue and black lines almost coincide since 88\% of the assessors are consistent in our study. However, other sensory 
tests may lead to substantially different ratio of consistent and inconsistent assessors, and the aggregated preferences vary depending on including or excluding the inconsistent group. Therefore such an analysis is recommended for each consumer sensory test.

Finally, the global liking score can be explained with the nine attributes of the experiment such that separate models are estimated for consistent and inconsistent assessors. The regression outputs can be found in the Appendix. Even though the regression models do not differ to a high extent, some differences appear as well:
\begin{itemize}[label=--]
    \item for consistent assessors, global taste plays a more robust role;
    \item for consistent assessors, all significant variables have positive coefficients, but this is not obvious for inconsistent assessors;
    \item a different set of variables is significant for the two groups;
    \item the explanatory power is better for consistent assessors (the values of the adjusted $R^2$ are 0.66 and 0.56 for the two groups, respectively).
\end{itemize}

\section{Discussion}

The results are discussed separately for each research question. Furthermore, we have some suggestions on how to conduct sensory tests in the future.

\subsection*{What type of metrics can be used to characterise consistent and inconsistent evaluation?}

To assess (in)consistency, at least two values should exist that can be compared. 

This approach requires the presence of two scales (liking, JAR), whose pairs of values provide the basis for the comparison. In our case, any assessor evaluated the same attribute in two different ways, first on a nine-category monotonic ascending hedonic response scale, and on a five-category just about right (JAR) scale.



\subsection*{How to rank assessors and attributes' evaluations based on (in)consistency?} 

The previous approach can be used, either to evaluate the attributes or to rank the assessors according to their level (in)consistency. The order of the attributes has a natural interpretation. For example, an inconsistent attribute may indicate that it is difficult to interpret or perceive by the consumers. Such an attribute should be specified or omitted from tests of the given product. In other words, consumer questionnaires need to take the ability, knowledge, and perception of consumers into account.

Assessors have different perceptual, cognitive, socio-cultural background. At the same time, the consumer test administered at a given time may be influenced by the individual's current health status, motivational state, task awareness, attention, concentration, hunger, mood, etc.   
If the inconsistency on a given trait affects a large number of raters, it indicates problems with the sensory trait. If it involves a small number of raters, then it is the individual's assessment that needs to be examined.

\subsection*{Who can be considered an (in)consistent assessor?}

If one has a ranking of assessors by their inconsistency, a simple threshold can be used to separate the set of assessors whose evaluations are probably unreliable for the given set. A natural choice is labelling an assessor inconsistent if their evaluations are seemingly random (or they have misunderstood the two scales), i.e., the associated Kendall $\tau_C$ is not significantly negative.

\subsection*{What is the difference between consistent and inconsistent assessors' evaluation and use of scale?}

In our dataset, the mean liking and JAR values of consistent raters and inconsistent raters do not differ significantly, but they do differ significantly in their standard deviations. The range of the scale is better used by the consistent assessors in both liking and JAR assessment. Further studies are needed to test this conjecture. Scale usage should also be analysed in any such test because it may be indicative of inconsistent evaluation.

\subsection*{What is the impact of consistent and inconsistent assessors' evaluations on the overall liking estimate?}

In our study, there is no significant difference between inconsistent and consistent assessors in the estimation of global liking. However, it is entirely possible that this effect varies from test to test, and significantly high effects may be observed for the evaluation of inconsistent assessors. Therefore, the impact of omitting inconsistent evaluations should be checked after each test.

\subsection*{Recommendations}

Using two different scales (e.g.\ liking and JAR) in sensory testing makes it possible to identify and measure the inconsistency of assessors. Nonetheless, it is important to ensure that they fully understand the two scales, which can be checked, for instance, by presenting two examples with Kendall $\tau_C$ values close to $-1$ and $0$ (such as in Table~\ref{Table4}) and asking which of them is more likely to be obtained from a past test. Furthermore, in an online questionnaire, the evaluations can be continuously monitored and the assessor might be warned if the last score is suspicious, e.g.\ an almost maximal liking score is followed by a JAR score of $-2$ or $+2$, when the assessor can be asked to review the most recent score. This technique can help to avoid typos and obvious mistakes.

Once the evaluations are collected, the process described in Section~\ref{Sec3} can be implemented:
\begin{enumerate}
    \item Kendall $\tau_C$ coefficients are calculated for all assessors;
    \item Assessors are called consistent and inconsistent according to whether the Kendall $\tau_C$ of their evaluations is significantly lower than zero (consistent) or not (inconsistent);
    \item The standard sample analysis is carried out for all consistent assessors, as well as for only the consistent assessors, and the results are compared.
\end{enumerate}
Generally, we think it is better to conclude only on the basis of consistent evaluations since inconsistent preferences may distort reality.

\section{Conclusions}

Consumer sensory testing is the basis for determining directions of product development in food industry practice. The reliability of consumer tests resides in the consistency of consumer assessment. In our study, the consumers evaluated different biscuit samples on two different scales (nine-category monotonic ascending hedonic response scale, five-category just about right (JAR) intensity scale), which allowed testing consistency.
Using these pairs of evaluations, Kendall $\tau_C$ values have been calculated to identify inconsistent evaluations, which might be considered with a lower weight or fully excluded from further assessment (e.g. preference mapping, or penalty analysis, etc.). A ranking of both assessors and attributes can be established, too. We found that consistent assessors are characterised by the use of a broader range of the liking scale and JAR scale. 

Further research is needed on the roots, factors and circumstances behind inconsistent evaluations: sensory fatigue, misunderstanding of scales, intentional manipulation, motivation, mental abilities, different perceptual abilities, product specificity, etc. For all consumer sensory tests having a similar design, it is advisable to test inconsistency since this is the only way to tell who has been affected by the inconsistent evaluation, for which sensory attribute and for which sensory question. Furthermore, future consumer tests are recommended to be adapt a design that makes it possible to identify inconsistent evaluations.

\section*{Acknowledgements}
\addcontentsline{toc}{section}{Acknowledgements}
\noindent
This research was funded by National Research, Development and Innovation Office,
OTKA, contracts number 135700.
\noindent

\bibliographystyle{apalike}
\bibliography{Bibliography}

\section*{Appendix}
\subsection*{Regression output for consistent assessors}
\begin{verbatim}
Coefficients:
                Estimate Std. Error t value Pr(>|t|)    
(Intercept)     -0.81294    0.21183  -3.838 0.000133 ***
colour           0.06173    0.02364   2.611 0.009184 ** 
global_odour     0.12827    0.03115   4.118 4.18e-05 ***
sweet_odour      0.04783    0.03091   1.547 0.122189    
margarine_odour -0.04080    0.03275  -1.246 0.213218    
global_taste     0.48071    0.03096  15.525  < 2e-16 ***
sweet_taste      0.10315    0.03257   3.167 0.001592 ** 
margarine_taste  0.11238    0.03105   3.619 0.000313 ***
hardness         0.16735    0.02058   8.132 1.45e-15 ***
crumbleness      0.08486    0.02344   3.621 0.000310 ***
---
Signif. codes:  0 ‘***’ 0.001 ‘**’ 0.01 ‘*’ 0.05 ‘.’ 0.1 ‘ ’ 1

Multiple R-squared:  0.6601,    Adjusted R-squared:  0.6565 
\end{verbatim}

\subsection*{Regression output for inconsistent assessors}
\begin{verbatim}
Coefficients:
                Estimate Std. Error t value Pr(>|t|)    
(Intercept)     -0.58162    0.67222  -0.865 0.388800    
colour          -0.07221    0.07742  -0.933 0.353032    
global_odour     0.18447    0.09435   1.955 0.053098 .  
sweet_odour      0.16118    0.09816   1.642 0.103442    
margarine_odour -0.17723    0.09003  -1.969 0.051517 .  
global_taste     0.29717    0.07766   3.827 0.000216 ***
sweet_taste      0.29706    0.08925   3.328 0.001188 ** 
margarine_taste  0.30966    0.07768   3.986 0.000121 ***
hardness         0.18614    0.06167   3.019 0.003157 ** 
crumbleness     -0.02762    0.06915  -0.399 0.690315    
---
Signif. codes:  0 ‘***’ 0.001 ‘**’ 0.01 ‘*’ 0.05 ‘.’ 0.1 ‘ ’ 1

Multiple R-squared:  0.5972,    Adjusted R-squared:  0.5642

\end{verbatim}

\subsection*{Regression output for all assessors}
\begin{verbatim}
Coefficients:
                Estimate Std. Error t value Pr(>|t|)    
(Intercept)     -0.73306    0.20278  -3.615 0.000316 ***
colour           0.04650    0.02272   2.047 0.040955 *  
global_odour     0.13847    0.02969   4.664 3.53e-06 ***
sweet_odour      0.06059    0.02963   2.045 0.041162 *  
margarine_odour -0.06306    0.03064  -2.058 0.039866 *  
global_taste     0.44907    0.02883  15.575  < 2e-16 ***
sweet_taste      0.12280    0.03066   4.006 6.65e-05 ***
margarine_taste  0.14075    0.02864   4.915 1.04e-06 ***
hardness         0.17121    0.01957   8.749  < 2e-16 ***
crumbleness      0.07089    0.02227   3.184 0.001498 ** 
---
Signif. codes:  0 ‘***’ 0.001 ‘**’ 0.01 ‘*’ 0.05 ‘.’ 0.1 ‘ ’ 1

Multiple R-squared:  0.6463,    Adjusted R-squared:  0.6431 

\end{verbatim}

\end{document}